\documentclass[aps, prl, amsmath,amssymb, reprint, showpacs,superscriptaddress,groupedaddress, nofootinbib ]{revtex4-1}

\usepackage{graphicx}
\usepackage{dcolumn}
\usepackage{bm}
\usepackage{amsmath}
\usepackage{color}

\usepackage{ulem}
\usepackage{dcolumn}  
\usepackage{amsmath}
\usepackage{multirow}
\usepackage{graphicx}   
\usepackage{subfigure}
\usepackage{comment}
\usepackage{color}
\usepackage[colorlinks,bookmarks=false,citecolor=blue,linkcolor=black,urlcolor=blue]{hyperref}
\usepackage{nicefrac}
\usepackage{soul}
\usepackage{slashbox}
\usepackage{bm,array}
\usepackage{tikz}
\usepackage{pifont}
\usepackage{gensymb}

\newcolumntype{C}{>{\centering\arraybackslash}p{1em}}

\begin{document}
	
	\title{Optoelectronic Properties and Defect Physics of Lead-free Photovoltaic Absorbers Cs$_2$Au$^{I}$Au$^{III}$X$_6$ (X=I, Br)} 
	\author{Jiban Kangsabanik,$^{1\dagger}$ Supriti Ghorui,$^{1\dagger}$ M. Aslam$^1$ and  Aftab Alam$^1$}
	\email{aftab@iitb.ac.in}	
	\affiliation{$^1$Department of Physics, Indian Institute of Technology Bombay, Powai, Mumbai 400076, India}
	\let\thefootnote\relax\footnote{$\dagger$ These two authors have contributed equally to this work}
	
	\begin{abstract}
		Stability and toxicity issues of hybrid lead iodide perovskite MAPbI$_3$ necessitate the hunt for potential alternatives.  Here, we shed new light on promising photovoltaic properties of gold mixed valence halide perovskites Cs$_2$Au$_2$X$_6$ (X=I, Br, Cl). They satisfy the fundamental requirements such as non-toxicity, better stability, band gap in visible range, low excitonic binding energy etc. Our study shows favorable electronic structure resulting in high optical transition strength, thus sharp rise in absorption spectra near band gap. This, in turn, yields very high short circuit current density and hence higher simulated efficiency  compared to MAPbI$_3$. However, careful investigation of defect physics reveals the possibility of deep level defects (such as V$_X$, V$_{Cs}$, X$_{Au}$, X$_{Cs}$, Au$_i$, Au$_{X}$, X= I, Br), depending on the growth condition. These can act as carrier traps and become detrimental to photovoltaic performance. The present study should help to take necessary precautions in synthesizing these compounds in a controlled chemical environment which can minimize the performance limiting defects and pave the way for future studies on this class of materials.																													
	\end{abstract}

	\maketitle
	
	\section*{I. Introduction}
	Since its inception in 2009, hybrid lead halide perovskite has become the center of attention in the photovoltaic community. High optical absorption and defect tolerance made its efficiency as high as 22.1\%, almost comparable to commercial silicon solar cells.\cite{kojima2009organometal, qian2016comprehensive, yin2015halide} Despite being highly efficient, it still has not been commercialized till date, mainly due to two reason, (i) poor stability in ambient environment (ii) toxicity due to Pb. While replacing the organic cation with inorganic Cs has helped in stability, but the presence of Pb seems insurmountable till now. A lot has been proposed as alternatives but either they were even more unstable or the efficiency is low.\cite{frolova2016exploring, kumar2016crystal, noel2014lead, yokoyama2016overcoming} One of the key alternatives which emerges recently are double perovskite halides, A$_2$BB$^{'}$X$_6$\cite{volonakis2016lead} where A is Cs, X is one of the halides, and B, B$^{'}$ are +1, +3 elements or vice versa. There exists various theoretical and experimental studies exploring different B, B$^{'}$ combinations, but most of these compounds either have indirect band gap (leading to higher recombination loss) or the gap is in high violet region due to the optically forbidden transition (leading to poor absorption).\cite{chakraborty2017rational, meng2017parity, xiao2017intrinsic} Some solution has been reported showing indirect to direct transition but toxicity was still a concern.\cite{slavney2017defect, tran2017designing, kangsabanik2018double} 

	Cs$_2$Au$_2$X$_6$ (X=I, Br, Cl) are a class of  compounds which show semiconducting behavior at ambient condition.\cite{liu1999electronic, kojima2000p} Although the predicted band gaps  for these materials are quite favorable for solar absorption, however they were  never being  investigated from the photovoltaic perspective. Very recently, Debbichi et al.\cite{debbichi2018mixed} reported a theoretical study on Cs$_2$Au$_2$I$_6$ and proposed it as a promising photovoltaic absorber. This was further confirmed by Giorgi et al.\cite{giorgi2018two} who showed the presence of weakly bound excitons in this compound, (similar to MAPbI$_3$) hinting towards good photovoltaic performance. However, few key points are not properly addressed in these studies e.g. (i) nature of band gap (ii) estimation of solar efficiency, etc. In addition, these studies are only focused on Cs$_2$Au$_2$I$_6$ compound, although the other halide compounds also have band gap in the visible range. Apart from these, defect physics of these compounds has never been studied. This is extremely important because defects in photovoltaic materials play a crucial role in dictating the device efficiency. For example, presence of a deep level defect which can act as electron-hole recombination center, limits the carrier diffusion to a greater extent.\cite{walsh2017instilling} While synthesizing, it is therefore very important to create a chemical environment which minimizes the defect concentration in a compound.

In this paper, we have performed a detailed first principle calculation to study the electronic, structural, and optical properties of the full series Cs$_2$Au$_2$X$_6$ (X=I, Br, Cl) from a perspective of photovoltaic applications. Careful analysis of band structure reveals slightly indirect nature of band gap, in contrast to earlier reports.\cite{debbichi2018mixed, giorgi2018two} However the optically allowed direct band gap remains within a few meV from the electronic gap, resulting in exciting optoelectronic behavior. Chemical, mechanical, and dynamical stability of all the compounds are also studied. In addition, we have investigated the possibility of point defect formation under various growth environments and found that even in anion rich condition, there is a possibility of the formation of deep level halide vacancies in Cs$_2$Au$_2$X$_6$. In addition, few other deep level defects are likely to form depending on the material and chemical environment, which may hinder its performance as photovoltaic absorber. Additionally, we have also simulated the series of compounds made of organic cations in place of Cs i.e. MA(CH$_3$NH$_3^{+}$), FA(CH(NH$_2$)$_2^+$), and investigated their possibility of formation and potential as photovoltaic absorber. All the calculations are done by employing first principles Density Functional Theory (DFT)\cite{kohn1965self} as implemented in Vienna Ab-initio Simulation Package (VASP).\cite{kresse1996efficiency,kresse1999ultrasoft}  Other details of the calculations are given in  Appendix A.

\begin{figure}[t]
	\centering
	\includegraphics[width=0.91\linewidth]{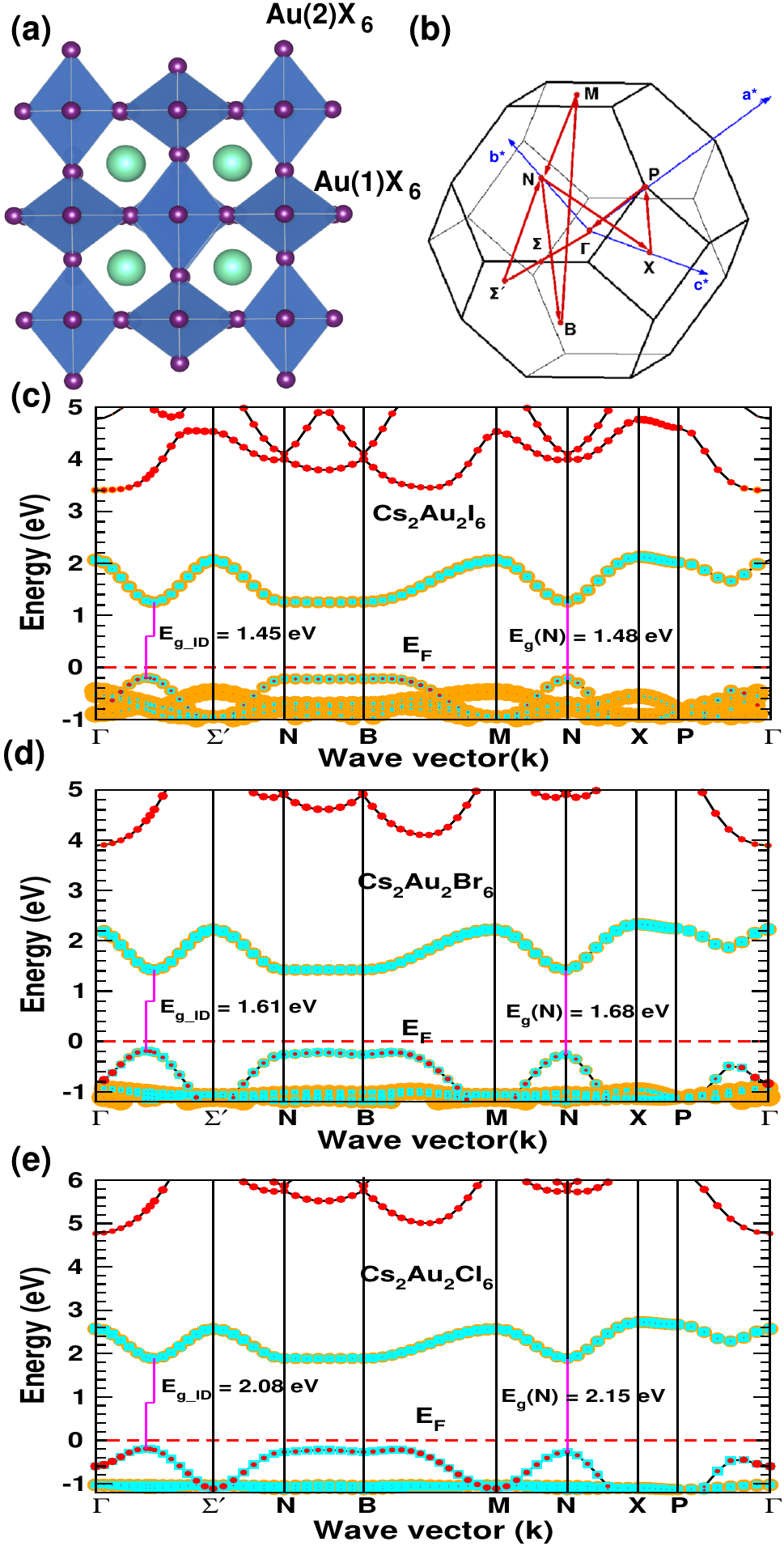}
	\caption{(a) Crystal structure and (b) 3D Brillouin zone of Cs$_2$Au$_2$X$_6$; X=I, Br, Cl. (c), (d), and (e) show HSE06 electronic band structure of Cs$_2$Au$_2$X$_6$; X=I, Br, Cl respectively, with band gap shifted to HSE06-G$_0$W$_0$ calculated values. Orange, turquoise, and red colored symbols indicate I-p, Au-d and Au-s orbital character.}
	\label{fig:1}
\end{figure}

\section*{II. Structural details and Stability}
Under the ambient condition, Cs$_2$Au$_2$X$_6$ (X=I, Br, Cl) crystallizes in double perovskite structure with space group I4/mmm( \#139).  In this structure, Cs atoms sit at 4d Wyckoff site, Au(1) and Au(2) at 2b and 2a respectively while the anions sit at two inequivalent 4e(X(1)) and 8h(X(2)) sites. Here Au(1) and Au(2) possess +1, and +3 oxidation state respectively, making it possible to have all the features of double perovskite. Halogens form alternate linear [Au(1)X$_2$]$^-$, and square-planar [Au(2)X$_4$]$^-$ complexes.\cite{kojima2000p, liu1999electronic} The  presence of alternately arranged elongated, and compressed AuI$_6$ octahedra can be seen in Fig. \ref{fig:1}(a), and can be confirmed from respective Au-X bond lengths (see SM\cite{supplement}), which matches well with previous experimentally reported data.\cite{riggs2012single}

First, we checked the chemical, mechanical, and dynamical stability of these compounds. For chemical stability, we have calculated the formation enthalpy ($\Delta H_f$) against the binary halides in the following pathway: $\text{M}_2\text{Au}_2\text{X}_6 \rightarrow 2\text{MX}+\text{AuX}+\text{AuX}_3$ (M=Cs, MA, FA; X=I, Br, Cl). They are presented in Table \ref{table:1} (for inorganic compounds). Going from I $\rightarrow$ Br $\rightarrow$ Cl, the chemical stability increases. For mechanical stability, we calculated the elastic constants (tabulated in SM\cite{supplement}) which satisfies the Born Huang mechanical stability criteria for all three halides.\cite{born1954dynamical} Calculated phonon dispersions (shown in SM\cite{supplement}) show no imaginary frequencies, and hence confirms the dynamical stability.

\begin{figure*}[t]
	\centering
	\includegraphics[width=1.0\linewidth]{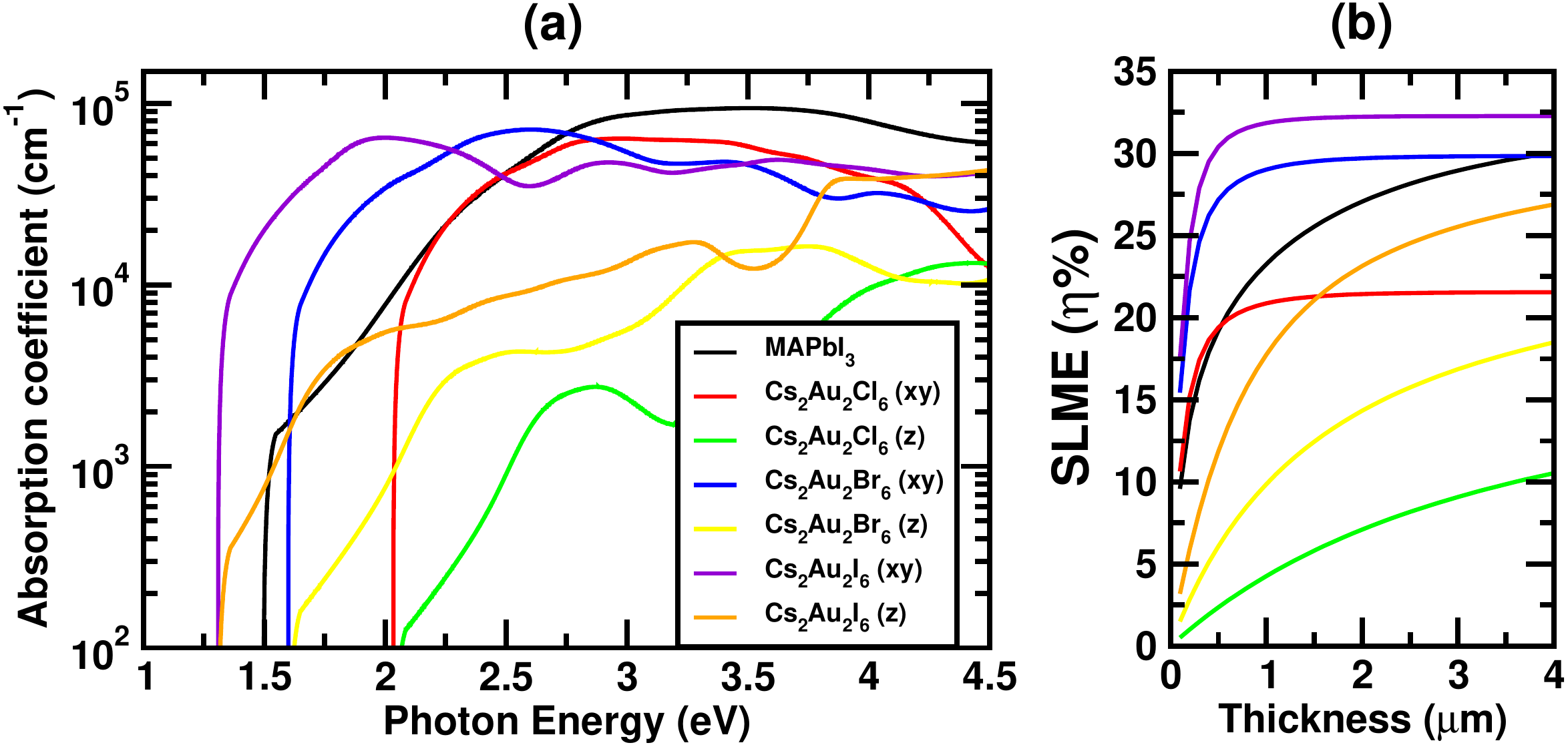}
	\caption{(a) Absorption coefficient vs. incident photon energy, and (b) Spectroscopic limited maximum efficiency (SLME) vs. film thickness at $298$ K for Cs$_2$Au$_2$X$_6$; X=I, Br, Cl, along xy-plane, and z-direction. For comparison,  simulated results for state of the art, MAPbI$_3$ is also plotted} 
	\label{fig:2}
\end{figure*}

\section*{III. Electronic structure}
Figure \ref{fig:1}(c, d, e) shows the electronic band structure calculated using hybrid (HSE06) functional\cite{krukau2006influence} for the three compounds. Cs$_2$Au$_2$I$_6$ forms an intermediate band comprised mainly of Au(2) $5d_{x^2-y^2}$ and I-p orbitals, which is responsible for its band gap (1.31 eV) in the visible range.\cite{kojima1994optical, giorgi2018two} The valence band maxima (VBM) consists mainly of Au(1) $5d_{z^2}$  and I-p orbital. In the case of Br and Cl, orbital contributions seem to be similar, but the band gap increases due  to increase in nearest neighbors Coulomb interaction and Jahn-Teller distortion.\cite{liu1999electronic}  For better accuracy, we have used HSE06 functional to simulate the band edge information, whereas, the band gap values are calculated using quasiparticle G$_0$W$_0$ calculations starting from wavefunction obtained using HSE06 functional. Our calculation reveals that all these materials have a slightly indirect band gap, having VBM and conduction band minima (CBM)  at different points along $\Gamma$ to $\Sigma^{'}$ direction, in contrast to the previous reports \cite{debbichi2018mixed, giorgi2018two}, where a direct band gap is predicted at high symmetry N-point. This is due to the fact that, a more detailed Brillouin zone sampling (i.e. considering an important $\Gamma$-$\Sigma^{'}$ direction in the band structure, where the actual VBM and CBM lies) is done in our study.  We have also  calculated the dipole transition matrix elements ({\it aka} transition probability) showing allowed optical transition at direct band gap [see Fig. S3 of SM\cite{supplement}]. High transition strength indicating the possibility of high absorption can be attributed to the mixing of halogen p and Au d orbitals.\cite{heo2017cutas3} Table \ref{table:1} shows our simulated band gap along with the difference between indirect and  optically allowed direct gap ($\Delta$$E_g^{da}$). Incorporation of spin orbit coupling (soc), does not show any significant effect on the optoelectronic properties of Cs$_2$Au$_2$I$_6$ (band gap (E$_g$) changes by 0.07 eV and $\Delta$E$_g^{da}$ remains same). This in turn, shows a negligible change in simulated efficiency. That is why, we have not used soc in our calculation for the other compounds, for which the effects are expected to be even less. Our simulated band gaps matches fairly well with previously reported experimental values.\cite{liu1999electronic} We have also checked the properties of organic cation namely MA and FA counterpart of these compounds. Our calculated lattice parameters for MA$_2$Au$_2$I$_6$ agrees well with previous experimental data.\cite{evans2017perovskite} All the other electronic structure data along with the band structure and transition probabilities are shown in SM.\cite{supplement} All the organic mixed valence gold perovskite compounds show fairly large band gap, restricting their application as photovoltaic absorber.
\begin{table*}[t]
	\begin{centering} 
		\resizebox{\textwidth}{!}{	\begin{tabular}{c c c c c c c c c c c }		
				\hline
				Compound & $\Delta H_f$ &$E_g$(eV)& $E_g$(eV) & $\Delta$$E_g^{da}$& J$_{sc}$ &J$_{max}$&V$_{oc}$ & V$_{max}$ & SLME & FF  \tabularnewline
				\vspace{0.1 in}  
				&  (meV/atom)& HSE06+GW& (Expt.)$^a$ & (meV) & (mA/cm$^2$) & (mA/cm$^2$) & (V) & (V) & $\eta$\% & \tabularnewline
				\hline 
				\vspace{0.05 in}  
				Cs$_2$Au$_2$I$_6$ &  -66.75&1.45& 1.31&13.3 & 33.02 & 32.15 & 1.04 & 0.95 & 30.41 & 0.89 \tabularnewline
				\vspace{0.05 in}  
				Cs$_2$Au$_2$Br$_6$   & -109.97&1.61 &1.60 &17.1 & 22.90 & 22.43 & 1.31 & 1.21 & 27.19 & 0.91 \tabularnewline
				\vspace{0.05 in}  
				Cs$_2$Au$_2$Cl$_6$ & -138.14 &2.08&2.04&16.2& 12.20 & 12.01 & 1.72 & 1.62 & 19.40 & 0.92  \tabularnewline
				\vspace{0.05 in}  
				MAPbI$_3$ & -71.65 & 1.72(SS-G$_0$W$_0$)$^b$& 1.50$^c$ & 0 & 16.76 & 16.40 & 1.27 & 1.17 & 19.21 & 0.90
				\\\hline  
		\end{tabular}}
		\par
	\end{centering}
	\caption{Formation enthalpy ($\Delta H_f$) simulated and experimental band gap ($E_g$), difference between electronic and optically allowed direct gap ($\Delta$$E_g^{da}$), short circuit current density (J$_{sc}$), open circuit voltage (V$_{oc}$), current density (J$_{max}$) and voltage (V$_{max}$) at maximum power, SLME and fill factor (FF) at 298 K for the three compounds. Various device related parameters are calculated at a film thickness of 500 nm. For comparison, relevant data for MAPbI$_3$ are also tabulated. $^{a}$\cite{liu1999electronic},  $^{b}$\cite{filip2014g},  $^{c}$\cite{yin2015halide}} 
	\label{table:1}
\end{table*}

	\section*{IV. Absorption coefficient and Spectroscopic Limited Maximum Efficiency (SLME)}	
Finite values of calculated transition probability encourage us to simulate the next relevant parameters for photovoltaic applications, i.e. absorption coefficients and spectroscopic limited maximum efficiency (SLME). Details about SLME formulation is given in SM,\cite{supplement} which is a better indicator of photovoltaic efficiency than the bare Shockley-Queisser limit.

Figure \ref{fig:2}(a) and \ref{fig:2}(b) show the absorption coefficients ($\alpha$) and SLME respectively for the three systems. For comparison, corresponding simulated data for MAPbI$_3$ are also shown. From Fig. \ref{fig:2}(a), one can see that the absorption coefficient in xy-plane is order of magnitude higher than those in the z-direction, confirming strong optical anisotropy of the material.\cite{debbichi2018mixed, giorgi2018two} This gives us an idea about appropriate alignment of the crystal axes, so as to maximize the photo absorption. Careful analysis reveals that the first optical peak along xy-direction can be attributed to vertical transition between two highest valence states (comprising of mainly Au(1) $5d_{z^2}$ orbitals) and two lowest conduction states (Au(2) $5d_{x^2-y^2}$ and I $p_z$ orbitals), whereas along z-direction it is due to the transition from lower part of the valence band (Au(1) $5d_{xz, yz}$) to the intermediate band, explaining the significantly higher optical absorption in xy-plane.\cite{giorgi2018two}

Looking at the absorption spectra one can see a sharp rise in absorption coefficient ($\alpha$) near the band gap for all three halides. Although in lower wavelength region the absorption coefficient is higher for MAPbI$_3$, but a lower band gap (for Cs$_2$Au$_2$I$_6$), and sharper rise of absorption spectra near band edge for both iodide and bromide compounds indicate better utilization of the solar spectrum. This can further be confirmed by our simulated short-circuit current density (J$_{sc}$).  For completeness, we have  tabulated room temperature (298 K) simulated values of few important device  parameters, such as J$_{sc}$, open-circuit voltage(V$_{oc}$), current density(J$_{max}$) and voltage(V$_{max}$) at maximum power output, SLME and fill factor(FF) at film thickness 500 nm for all three halides in Table \ref{table:1}, and compared the same with MAPbI$_3$. Notice that, J$_{sc}$ and J$_{max}$ are almost twice for Cs$_2$Au$_2$I$_6$ as compared to MAPbI$_3$, which becomes almost 1.5 times at saturation thickness. Slightly lower attainable voltage makes the efficiency 1.5 times at lower film thicknesses, and at least 3\% higher at saturation thickness. A little higher band gap makes Cs$_2$Au$_2$Br$_6$ to have higher attainable voltage making the SLME much higher than MAPbI$_3$ at lower film thicknesses, while it becomes comparable when thickness goes near saturation (see Fig. \ref{fig:2}).
For Cl, efficiency remains much lower compared to the other two halides, mainly due to higher band gap resulting in much lower attainable current.

\begin{figure*}[t]
	\centering
	\includegraphics[width=0.95\linewidth]{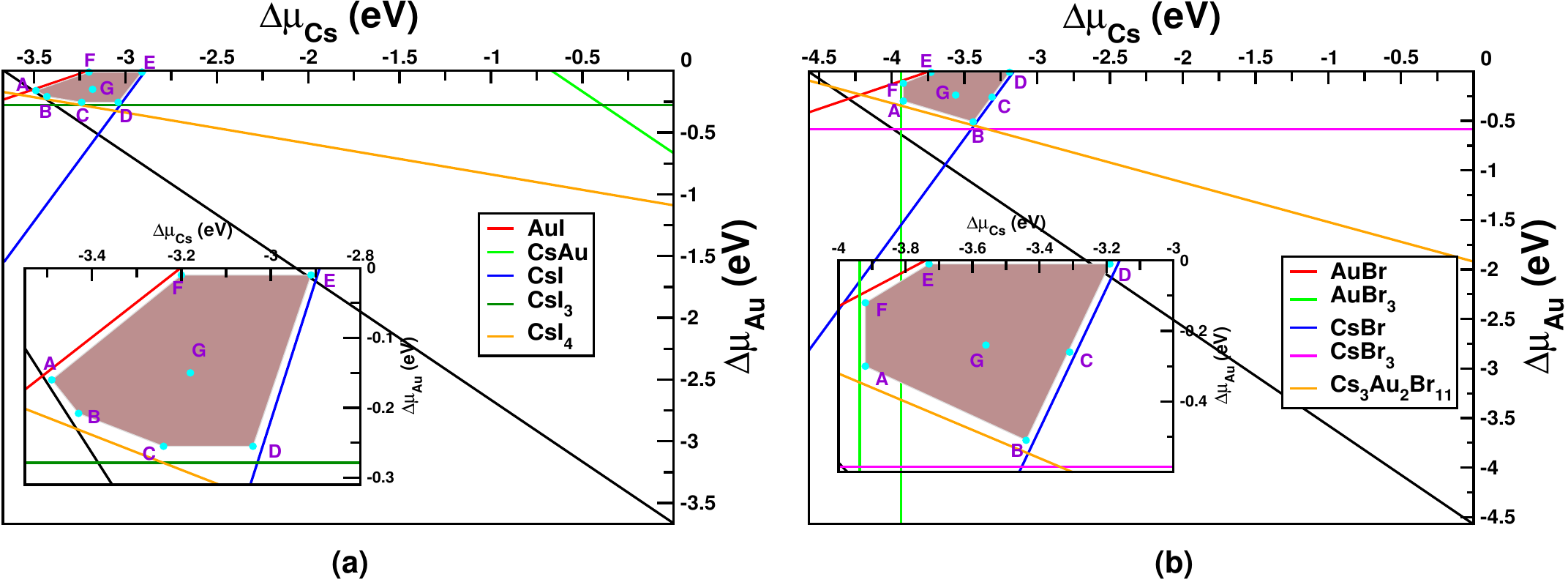}
	\caption{Phase diagram (allowed chemical potential region for constituents)  for  (a) Cs$_2$Au$_2$I$_6$ and (b) Cs$_2$Au$_2$Br$_6$, with respect to competing elemental, binary and ternary compounds. Brown shaded regions show the stable region for these double perovskite halides (zoomed view is given in the inset). Stable regions for the competing phases are shown by different color lines and their covered spaces in the opposite direction of the shaded region.   } 
	\label{fig:3}
\end{figure*}	

\begin{figure*}[t!]
	\centering
	\includegraphics[width=1.01\linewidth]{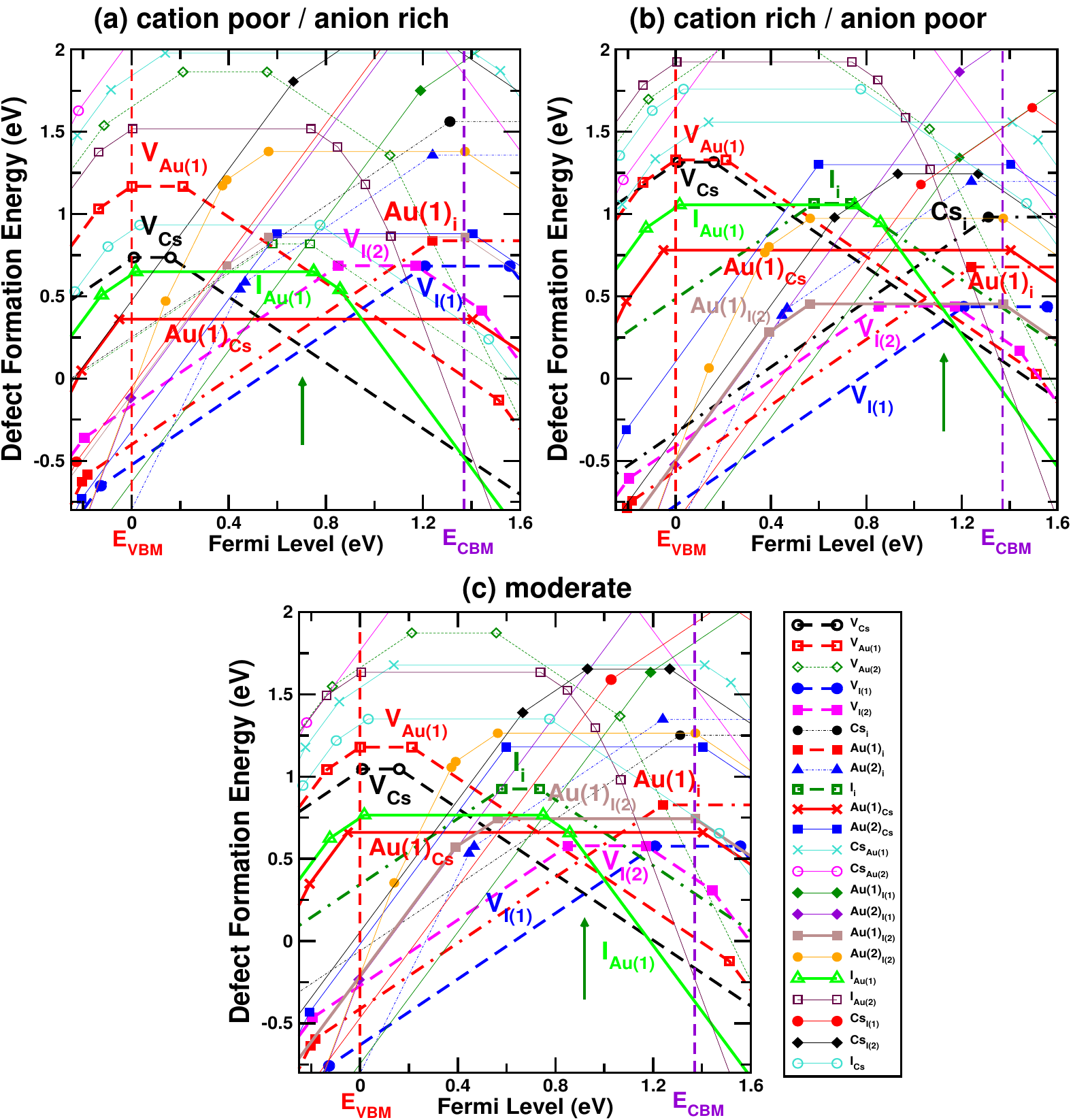}
	\caption{Defect formation energy as a function of Fermi level (E$_\text{F}$) for Cs$_2$Au$_2$I$_6$, for three different growth conditions (a) cation poor / anion rich and (b) cation rich / anion poor and (c) moderate cation / anion. In the phase diagram these conditions are represented by $`$A', $`$E', $`$G', points respectively. According to eq. 1, the charge state q of the defect is denoted by the slope of the function and the Fermi level at the turning point gives the charge transition energy level. Here each charge transition point for donor (acceptor) defects are indicated via filled (hollow) points. Region left to E$_{VBM}$ and right to E$_{CBM}$, presents the valence band below VBM and conduction band above CBM respectively. Green arrows point toward the Fermi level pinning} 
	\label{fig:4}
\end{figure*}

\begin{figure*}[t!]
	\centering
	\includegraphics[width=1\linewidth]{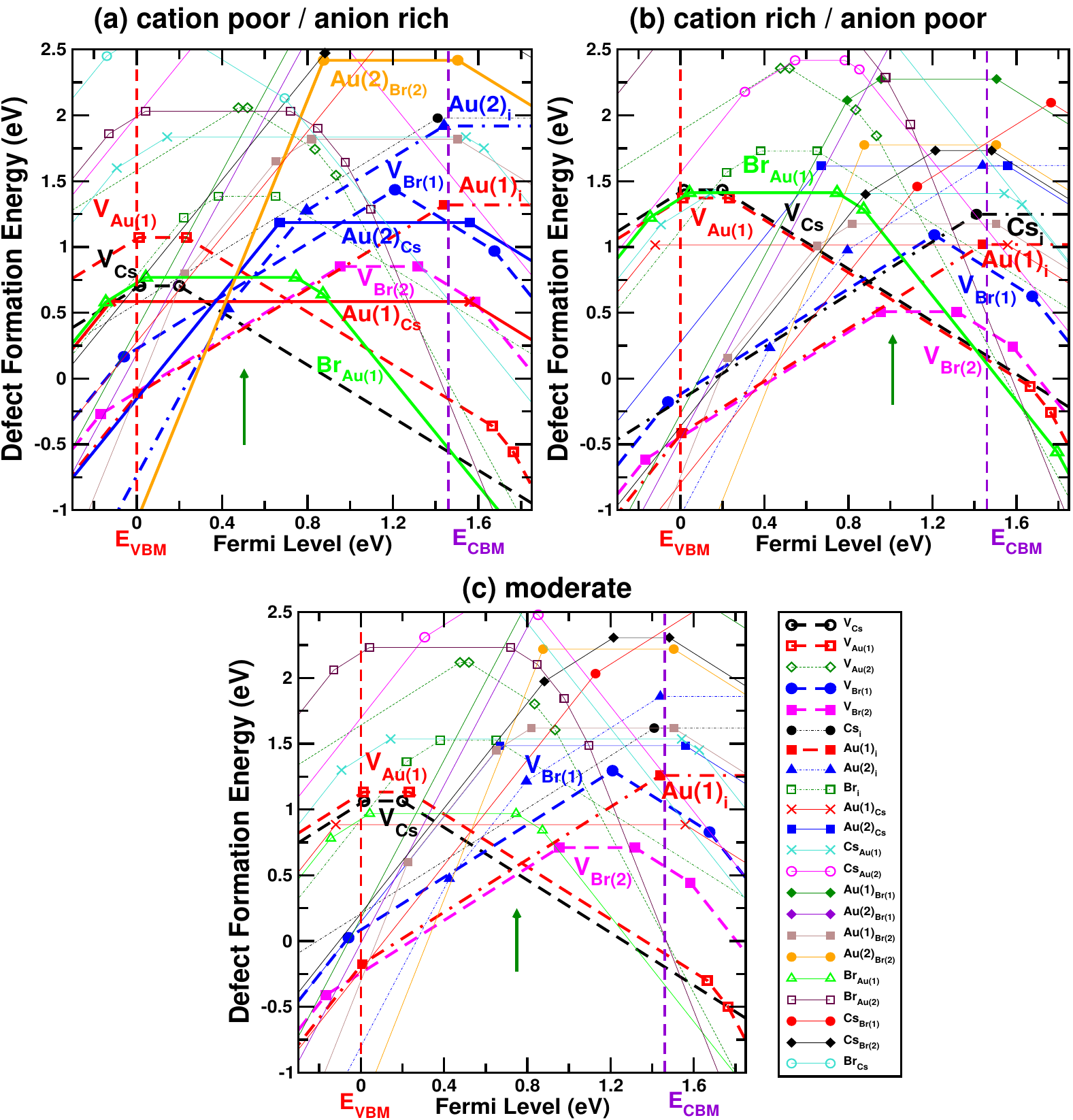}
	\caption{Defect formation energy as a function of Fermi level (E$_\text{F}$) for Cs$_2$Au$_2$Br$_6$, for three different growth conditions (a) cation poor / anion rich and (b) cation rich / anion poor and (c) moderate cation / anion. In the phase diagram these conditions are represented by $`$A', $`$D', $`$G', points respectively. According to eq. 1, the charge state q of the defect is denoted by the slope of the function and the Fermi level at the turning point gives the charge transition energy level. Here each charge transition point for donor (acceptor) defects are indicated via filled (hollow) points. Region left to E$_{VBM}$ and right to E$_{CBM}$, presents the valence band below VBM and conduction band above CBM respectively.  Green arrows point toward the Fermi level pinning} 
	\label{fig:5}
\end{figure*}

\section*{V. Defect Physics}
Defects play a major role in dictating the carrier mobility, lifetime, and recombination rate for a given semiconductor. Unlike extended defects (e.g. grain boundaries, surface passivation, etc.), intrinsic point defects (vacancies, interstitial, antisites, etc.) are very difficult to control.\cite{wallace2017steady, walsh2017instilling, Park2018} For example, in case of MAPbI$_3$, shallow dominant point defects\cite{yin2014unusual}  result in high carrier diffusion length aiding to its high efficiency. Whereas presence of deep level defects, acting as Shockley-Read-Hall (SRH) recombination centers, are known to be one of the main reasons behind significantly lower efficiency in case of kesterite (CZTS) solar cells.\cite{wallace2017steady} In order to gain  better insight, we performed a detailed ab-initio study of all the possible point defects in Cs$_2$Au$_2$I$_6$ and Cs$_2$Au$_2$Br$_6$, which are predicted to have comparable or higher theoretical efficiency than MAPbI$_3$.

Formation energy for a defect $D$ at a charge state `$q$' is defined as,
\begin{eqnarray}
E^{form}[D^q]=E^{tot}[D^q] & -E^{tot}[bulk]-\sum_{i}^{}n_i\mu_i + \nonumber \\
&	q(E_{VBM}+ \Delta E_F)+E_c\
\end{eqnarray}

Where $E^{tot}$[$D^q$] is the total energy for a supercell with the associated defect $D$ at a charge state `$q$'. $E^{tot}[bulk]$ is the total energy for pure bulk supercell of equivalent size. $\mu$$_i$ is the chemical potential of the associated defect with $n_i$ being the number of defects added ($n_i>0$) or removed ($n_i<0$). The next term accounts for the chemical potential for electrons added ($q<$0) or removed ($q>$0) to create various charged defect states. $E_{VBM}$ is the energy at valence band maxima, $\Delta E_F $ can be varied from 0 (at VBM) to band gap $E_g$ (at CBM). $E_c$ is the correction term which includes electrostatic and potential alignment corrections for charged defects.\cite{freysoldt2009fully} We also include the correction for band gap underestimation by PBE exchange correlation functional,\cite{perdew1996generalized} via incorporating the band edge (both VBM and CBM) shifts obtained by more accurate quasiparticle G$_{0}$W$_{0}$ calculations.\cite{lany2008assessment} More details about band gap corrections are provided in Appendix B.   

Three types of defects are considered, vacancies (V$_{Cs}$, V$_{Au}$, V$_{X}$), interstitials (Cs$_i$, Au$_i$, X$_i$), and anti-sites (Cs$_{Au}$, Cs$_{X}$, Au$_{Cs}$, Au$_{X}$, X$_{Cs}$, X$_{Au}$), (X=I, Br), etc. Two inequivalent Wyckoff positions for both  X and Au are considered. To accurately calculate various defect charge state energies, a 160 atom supercell is used. Further discussions on the choice of defects are presented in Appendix C. 

It is extremely important to notice that, $E^{form}(D^q)$ can vary depending on the particular choice of $\mu$.  Experimentally, this chemical potential can vary depending on the growth environment. The choice of $\mu$ generally depends on the stability of the compound against possible elemental and/or competing secondary phases. As secondary phases, we have considered the most stable binary halides and other super-ordered structures of the cations.

First, for the compound to be stable the below thermodynamic equilibrium must be reached,
\begin{equation}
2\Delta\mu_{Cs}+2\Delta\mu_{Au}+6\Delta\mu_X=\Delta H_f (X=Br,I)
\end{equation}
Here, $\Delta H_f$ is the formation enthalpy of the compound against its elemental constituents.  $\Delta\mu_i=\mu_i-\mu^0_i$ where $\mu^0_i$ is total energy of constituent `$i$' at its elemental phase. Following are a set of equations, which should be satisfied to avoid co-existence of elemental  and secondary phases, 
\begin{equation}
\Delta\mu_i < 0, {i=Cs,Au,X}
\end{equation}
\begin{equation}
a\Delta\mu_{Cs}+b\Delta\mu_{Au}+c\Delta\mu_X < \Delta H_f (Cs_aAu_bX_c)
\end{equation}
where, a,b,c=0,1,2....Z.

Figure \ref{fig:3}(a,b) shows the phase diagrams for Cs$_2$Au$_2$I$_6$ and Cs$_2$Au$_2$Br$_6$  against their possible competing elemental and secondary compound phases (taken from Materials Project database.\cite{jain2013commentary}) The brown shaded areas in both the figures show allowed chemical potential region for the constituents, keeping the intended material stable. We have taken seven different sets of allowed chemical potentials for each halide (shown as A,B,C,D,E,F,G points in respective diagrams), which represent various chemical environments, from cation poor/anion rich to cation rich/anion poor conditions. We have drawn the respective defect formation energy diagram for each of them. The related $\Delta\mu_i$ values and corresponding defect formation energy plots can be found in SM.\cite{supplement}

Figures \ref{fig:4} and \ref{fig:5} show the defect formation energies for various intrinsic defects as a function of Fermi level in three different chemical growth conditions, (a) cation poor/anion rich and (b) cation rich/anion poor, and (c) moderate cation/anion, for Cs$_2$Au$_2$I$_6$ and Cs$_2$Au$_2$Br$_6$ respectively. These three phases are marked as point $`$A', $`$E' ($`$D' for X=Br) and $`$G' in the respective phase diagrams in Fig. \ref{fig:3}, and corresponding $\Delta\mu$ values can be found in SM.\cite{supplement}  Most probable defects (setting a cut-off of E$^{form}$[D$^q$] $\sim$0.75 eV) are highlighted with respect to the Fermi level pinning in these figures. We have also considered a few other possible growth environments (points $`$A'-$`$G' in Fig. \ref{fig:3}), whose defect formation energies are shown in SM.\cite{supplement} Line styles (symbols) for various defects, in all the three growth environments, are shown in the right hand panel of Fig. \ref{fig:4}(c) and \ref{fig:5}(c).  For Cs$_2$Au$_2$I$_6$, at cation poor/anion rich environment (Fig. \ref{fig:4}(a)), the dominant acceptor type defects are V$_{Cs}$, V$_{Au(1)}$, (I$_{Au(1)}$ will act as neutral defect at this Fermi level pinning) whereas V$_I$ and Au(1)$_i$ are dominant donors. There is another neutral defect Au(1)$_{Cs}$, which can form at a high concentration, but because it does not show any charge state transition in the band gap, the Fermi level will be pinned at almost middle of the gap where V$_{Cs}$ and V$_{I(1)}$ intersects. At cation rich/anion poor environment (Fig. \ref{fig:4}(b)), donor defects (mostly iodine vacancies) are likely to drag the Fermi level more close to the CBM, where the main acceptor defects will be cation (Cs, Au(1)) vacancies and I$_{Au(1)}$. At this growth condition, the material will show low n-type carrier concentration. At moderate cation/anion condition (Fig. \ref{fig:4}(c)), the Fermi level pinning will be at a position near the mid gap region, indicating very low intrinsic carrier concentration. Overall, one can see that iodine vacancies (along with V$_{Cs}$) are most dominant ones at all growth conditions, in which V$_{I(1)}$ is more probable than V$_{I(2)}$. Apart from these, there are other possible defects which are likely to dominate depending on the chemical environment. These defects show deep transition levels well in the band gap, and may act as SRH recombination centers, hindering the photovoltaic (PV) efficiency. In case of bromide compound, vacancy at two inequivalent halide sites show different formation energies, with V$_{Br(2)}$ being the more probable one. At cation poor/anion rich environment (Fig. \ref{fig:5}(a)), contribution from acceptor V$_{Cs}$ will be mainly compensated by donors V$_{Br(2)}$ and Au(1)$_i$, pinning the Fermi level at the p-type side of the band gap. At this Fermi level pinning, a number of other deep level donor defects (such as Au(2)$_{Cs}$, Au(2)$_{Br(2)}$, Au(2)$_{i}$, V$_{Br(1)}$, etc.) and acceptor V$_{Au(1)}$ is likely to form. A very low n-type concentration is expected at cation rich/anion poor growth condition (see Fig. \ref{fig:5}(b)). Most of the defects will still be present when a moderate cation/anion environment is maintained (Fig. \ref{fig:5}(c)), pinning the Fermi level in the middle of the band gap. In all these growth environments, Fermi level pinning is mostly positioned near the mid gap region, which explains the experimental observation of high resistivity in these compounds. \cite{riggs2012single}

Overall analysis of these two halides from stability (phase diagram) and defect tolerance perspective indicates that, utmost care needs to be taken while synthesizing these compounds because the single phase stability region for both of them are rather small.
 In addition, both the compounds show possibility of numerous deep level defects (even more so given the high temperature ($\sim$900 K) synthesis procedure reported in \cite{riggs2012single}) which may act as carrier traps and thus substantially reduce the V$_{OC}$, in practice. As such, in spite of the excellent optoelectronic properties, Cs$_2$Au$_2$X$_6$ may have limited PV performance due to the possibility of deep level defects. Nevertheless, there exists few well known compounds, such as CIGS, CZTS etc., where the formation of defect complexes has been reported to make the deep level defects electrically benign.\cite{Park2018} In case of gold mixed halides, a number of defect complexes can be possible, thus requiring further studies to get a deeper understanding.

\section*{VI. Conclusion}

In summary, we have carried out a detailed analysis of optoelectronic properties and defect physics of Cs$_2$Au$_2$X$_6$ (X=I, Br), two compounds with several favorable properties such as non-toxicity, better stability and simulated efficiency  etc. First principles simulations predict these compounds to have slightly indirect nature of band gap, with optically allowed direct band gap remaining very close (within 20 meV) to the indirect gap, allowing the optical absorption to be very high. The value of band gap falling in the visible region and sharp rise of absorption coefficient near band edge yields reasonably high simulated efficiency (even at very small film thickness). Our study on defect physics, however, predicted the possibility of deep level defects in both of these materials. Halide vacancies are observed to be most dominant defects. Cation vacancies (V$_{Cs}$, V$_{Au(1)}$), interstitials and few antisite defects will also form depending on the material and growth environment. The existence of deep level defects (which can act as trap states) will probably make the  materials prone to carrier loss due to non-radiative recombination. Analysis of the defect physics (Fermi level pinning, dominant defects, etc.) explains some of the experimental observations reported earlier. Nevertheless, we believe that, the present study will guide experimentalists to employ optimal chemical growth conditions to carry out future studies on these compounds and also help theoreticians to work on similar aspects of related materials.\cite{gajdovs2006linear, yu2012identification, shockley1961detailed, green2004third, tiedje1984limiting} 

\section*{Acknowledgements}
JK and SG have contributed equally to this work. They acknowledge financial support from IIT Bombay for research fellowship. AA and MA  acknowledge National Center for Photovoltaic Research and Education (NCPRE) funded by Ministry of new renewable energy (MNRE), Government of India and IIT Bombay for possible funding to support this research.

\section{ APPENDIX A: Computational Details  }	
All the calculations are done using Density Functional Theory (DFT) \cite{kohn1965self} as implemented in Vienna Ab-initio Simulation Package (VASP) \cite{kresse1996efficiency, kresse1999ultrasoft} with projector augmented Wave (PAW) basis set. Finding the equilibrium structure, calculation of  decomposition enthalpy and other primary electronic structure (band structure, density of states etc.) were done using Perdew-Burke-Ernzerhof (PBE) exchange correlation functional.\cite{perdew1996generalized}  Cs (5s$^2$5p$^6$6s$^1$), Au (5d$^{10}$6s$^1$), I (5s$^2$5p$^5$), Br (4s$^2$4p$^5$), Cl (3s$^2$3p$^5$), C (2s$^2$2p$^2$), N (2s$^2$2p$^3$), and H (1s$^1$) are used as valence electrons. An energy cutoff 500 eV with 6$\times$6$\times$6 $\Gamma$ centered k-mesh  is considered for structural optimization and unit cells are relaxed until forces reached to the value less than 0.001 eV/\r{A} for Cs$_2$Au$_2$X$_6$; X=I, Br, Cl. For organic gold halides, we have replaced the Cs with MA (CH$_3$NH$_3$) and FA (CH(NH$_2$)$_2$) cations in the relaxed Cs$_2$Au$_2$I$_6$ structure and then relaxed in 3 steps. At first,  we did full geometrical relaxation  then again performed relaxation with parameters keeping volume and shape fixed  and at last we did full geometrical relaxation  with 500 eV with 6$\times$6$\times$6 kpoints until forces converge to 0.01 eV/\r{A}. Charge densities were calculated using energy cutoff of 450 eV,  8$\times$8$\times$8 k-mesh using the relaxed structures until energy converges up to 10$^{-6}$ eV. Next, we have used Heyd-Scuseria-Ernzerhof (HSE06)\cite{krukau2006influence} functional  to get the band edge information. To obtain  more accurate value of band gap we have used G$_0$W$_0$ method along with HSE06 and PBE exchange correlation functional. Optical absorption coefficients are calculated within the independent particle approximation with PBE exchange correlation functional and then scissor shifted to experimental band gap while calculating the SLME. For simulation of various defects at different charge states, we have used 520 eV plane wave energy cutoff along with gamma centered k-mesh. For each defect in different charge states, we have only relaxed the ionic positions keeping the cell shape and volume fixed.

\subsection*{APPENDIX B: Corrections associated with charged defects and Band gap underestimation }
For charged defects, there are a few sources of error that comes from the DFT approximation which uses a background charge to neutralize the supercell. This requires two corrections, one is electrostatic interaction term correction and the other is potential alignment term. We use the well documented Freysoldt, Neugebauer and Van de Walle (FNV) scheme to correct these errors.\cite{freysoldt2009fully} Next we notice that for both the halides, PBE underestimates the band gap substantially. In practice, this can be seen as the shift of VBM (CBM) up (down) by $\Delta E_{VBM}$ ($\Delta E_{CBM}$) while showing a reduced band gap. This induces an underestimation in the defect formation energy by $\Delta E_{VBM}$ ($\Delta E_{CBM}$) per hole (electron) occupying the acceptor (donor) level, in case of acceptor (donor) type defects. Here to correct these, we calculate the shift in band edges from quasiparticle G$_{0}$W$_{0}$ calculations, which reproduce the experimental band gap very well in case of these two halides.\cite{lany2008assessment} We calculate the $\Delta E_{VBM}$ ($\Delta E_{CBM}$) to be 0.39 eV (0.26 eV) in case of Cs$_2$Au$_2$I$_6$ and 0.49 eV (0.30 eV) for Cs$_2$Au$_2$Br$_6$. All the defect formation energy calculations include the above mentioned corrections.

Note:  We have used the same set of pseudopotentials for all the defect related calculations including phase diagrams, band edge shifts. 

\subsection*{APPENDIX C: Choice of defects } 

Three types of defects are considered, vacancies (V$_{Cs}$, V$_{Au}$, V$_{X}$), interstitials (Cs$_i$, Au$_i$, X$_i$), and anti-sites (Cs$_{Au}$, Cs$_{X}$, Au$_{Cs}$, Au$_{X}$, X$_{Cs}$, X$_{Au}$), (X=I,Br), etc.. For these compounds, there are two inequivalent Wyckoff positions for both halides, X and Au, which we have considered while considering the vacancy and anti-site defects. For the interstitials, we have considered all the possible positions and chosen the final position based on the total energy calculated at neutral charge state. As discussed earlier we denote Au$^{+1}$ and Au$^{+3}$ as Au(1) and Au(2) respectively. X(1) and X(2) (X=Br,I) represents halide anions at 4e and 8h Wyckoff sites respectively.  Here we have used PYCDT code\cite{broberg2018pycdt}  to generate the supercells with defects. To accurately calculate the various defect charge state energies a 160 atom supercell is used.

\end{document}